\documentclass[10pt,aps,prb, twocolumn,superscriptaddress,amsmath,amssymb]{revtex4-1}

\usepackage[english]{babel}

\usepackage{comment}

\usepackage{graphicx}
\usepackage[colorlinks=true, allcolors=blue]{hyperref}

\newcommand\affiltheos{Theory and Simulation of Materials (THEOS) and National Centre for Computational Design and Discovery of Novel Materials (MARVEL), 
    \'Ecole Polytechnique F\'ed\'erale de Lausanne,
    1015 Lausanne, Switzerland}
\newcommand\affildtu{Department of Energy Conversion and Storage, Technical University of Denmark, DK 2800 Kgs. Lyngby, Denmark}
\newcommand\affilcapex{CAPeX Pioneer Center for Accelerating P2X Materials Discovery, DK 2800 Kgs.
Lyngby, Denmark}
\newcommand\affilpsi{PSI Center for Scientific Computing, Theory and Data, Paul Scherrer Institute, 5232 Villigen PSI, Switzerland}

\begin{document}
\title{Implementing a Scalable, Redeployable and Multitiered Repository for FAIR and Secure Scientific Data Sharing: The BIG-MAP Archive}

\author{Valeria Granata}
\thanks{VB and FL contributed equally.}
\affiliation{\affiltheos}
\author{Fran\c{c}ois Liot}
\thanks{VB and FL contributed equally.}
\affiliation{\affiltheos}
\author{Xing Wang}
\affiliation{\affilpsi}
\author{Steen Lysgaard}
\affiliation{\affildtu}
\author{Ivano E. Castelli}
\affiliation{\affildtu}
\author{Tejs Vegge}
\email{teve@dtu.dk}
\affiliation{\affildtu}
\affiliation{\affilcapex}
\author{Nicola Marzari} 
\affiliation{\affiltheos}
\affiliation{\affilpsi}
\author{Giovanni Pizzi}
\email{giovanni.pizzi@psi.ch}
\affiliation{\affilpsi}
\date{\today}

\begin{abstract}
Data sharing in large consortia, such as large research collaborations or industry partnerships, often requires addressing both organizational and technical issues. A common platform for data sharing is essential for the successful operation of the consortium, promoting collaboration within its projects, and facilitating the exchange of research findings. Among the several technical challenges of developing such a platform, like designing a scalable and redeployable architecture and creating a user-friendly interface for uploading and downloading data, security and access control are the primary concerns that need to be addressed to prevent unauthorized access or misuse of sensitive data.
Here, we describe the BIG-MAP Archive, a cloud-based, disciplinary, and private repository designed to address these challenges. Built upon the InvenioRDM platform, the repository leverages some of the functionalities of the platform to meet the specific needs of a consortium and provides a more adapted solution compared to general data repositories. Access to the repository can be restricted to members of a specific community, like the BATTERY 2030+ initiative for which it was originally built: a consortium of projects (among which BIG-MAP) aimed at accelerating the development and deployment of advanced battery technologies. Once data and metadata are uploaded to the Archive, access is restricted (with fine-grained access control, allowing access only to members of a single BATTERY 2030+ project, or to all initiative members). Furthermore, the formalized upload process ensures that data and metadata are already formatted and ready for publication in an open repository, when needed (e.g., upon publication of the corresponding scientific paper).
This paper reviews the key features of the repository, highlighting how the BIG-MAP Archive overcomes the challenges of sharing data within a large consortium by providing a secure and controlled environment for data storage and access, ensuring data confidentiality while enabling a flexible permissions-based access system, and how it can be easily redeployed to cover the needs of other consortia beyond BATTERY 2030+, such as the new MaterialsCommons4.eu and RAISE (Resource for AI Science in Europe) initiatives in the EU.
\end{abstract}
\maketitle

\section{Introduction}
Digital repositories serve to store and preserve data and metadata, facilitating access to research information and enabling scientists to share, reuse, and build upon the findings of each other. They have become essential tools for the scientific community and their number is continuously rising\cite{repos_materials,Pampel2013,Pampel2023}, often driven by funding agencies that require data related to publications to be published in appropriate public repositories\cite{NIH-opendata,SNF-opendata}.

Public data repositories are typically designed to share data openly and make it accessible to anyone, which may not be desirable or appropriate for collaborations that treat sensitive or confidential information. Situations may arise when researchers want to share their research data immediately as soon as it is produced, but only with certain colleagues before making it public (e.g., when the research is published in a scientific journal). This might happen, for example, as part of a federated self-driving autonomous laboratory campaign, driven by tools like FINALES\cite{finales,finales2}. Or they may need feedback from colleagues, be waiting for the approval of a license or patent, or have research that is part of a large consortium involving partnerships with companies and industries and that contains confidential data or information.

These challenges become central in large research consortia, which act as common frameworks for coordinating collaborative research efforts across multiple institutions. These consortia bring together experts from diverse disciplines to tackle ambitious goals, usually involving large amounts of data. The sharing of data within a consortium is essential for researchers to access the necessary information to advance in their work. Collaborative data repositories are therefore critical tools, serving as a hub for sharing and preserving data across multiple projects.
When consortia include partnerships with companies and industries, there is an even greater need to balance collaboration with security. In such cases, restricted data repositories offer enhanced control over resources compared to public repositories. By limiting access to data, these repositories enable consortia to maintain the confidentiality and integrity of the information while still promoting collaboration among partners.

To meet the needs of a large consortium, the repository should implement mechanisms to regulate access through authentication and authorization systems, and permission settings to provide fine-grained access control, allowing users to grant or deny access to specific datasets. Furthermore, transparency and confidentiality should be maintained by tracking access, user actions, and data usage, where appropriate. 

Furthermore, along with storing data and ensuring long-term data preservation, a data repository should make data easily accessible and discoverable so that it can be efficiently reused by others. To improve discoverability, the repository should provide or encourage standardized formats for data and metadata. Standardized formats are also needed to make data and metadata interoperable with other datasets, or data processing and analysis tools. This is nowadays an even more stringent requirement when considering making (FAIR) data available to Artificial Intelligence (AI) systems and machine learning (ML) tools.
Standardized formats can help facilitating sharing and reusing of data across different AI or ML applications by providing a common language and structure for describing data. This, in turn, enables AI systems to effectively learn from the aggregated data, leading to more accurate predictions.

Besides general data standards such as JSON (JavaScript Object Notation) and XML (Extensible Markup Language), and metadata standards like Dublin Core\cite{dublincore} and DataCite\cite{datacite}, more sophisticated techniques based on ontologies and Semantic Web are emerging, to help scientists more easily find the relevant data they need.
Ontologies and Semantic Web Technologies enable computers to understand and process data by creating languages that describe complex relationships between data\cite{Clark2025}, allowing for the retrieval and management of information based on meaning and logical connections. This means that computers can search for content not only by matching words, but also by connecting related concepts, entities, and relationships to provide more accurate and relevant search results. Researchers have explored the application of ontologies and Semantic Web Technologies in scientific fields.
Specialized standards are emerging to address specific domains, an example being the Battery Interface Ontology (BattINFO) \cite{battinfo}, an ontology of batteries and their interfaces, with domain-specific standards designed to formalize the current state of knowledge on battery interfaces.

\begin{figure}[tb]
\centering
\fbox{\includegraphics[width=\linewidth]{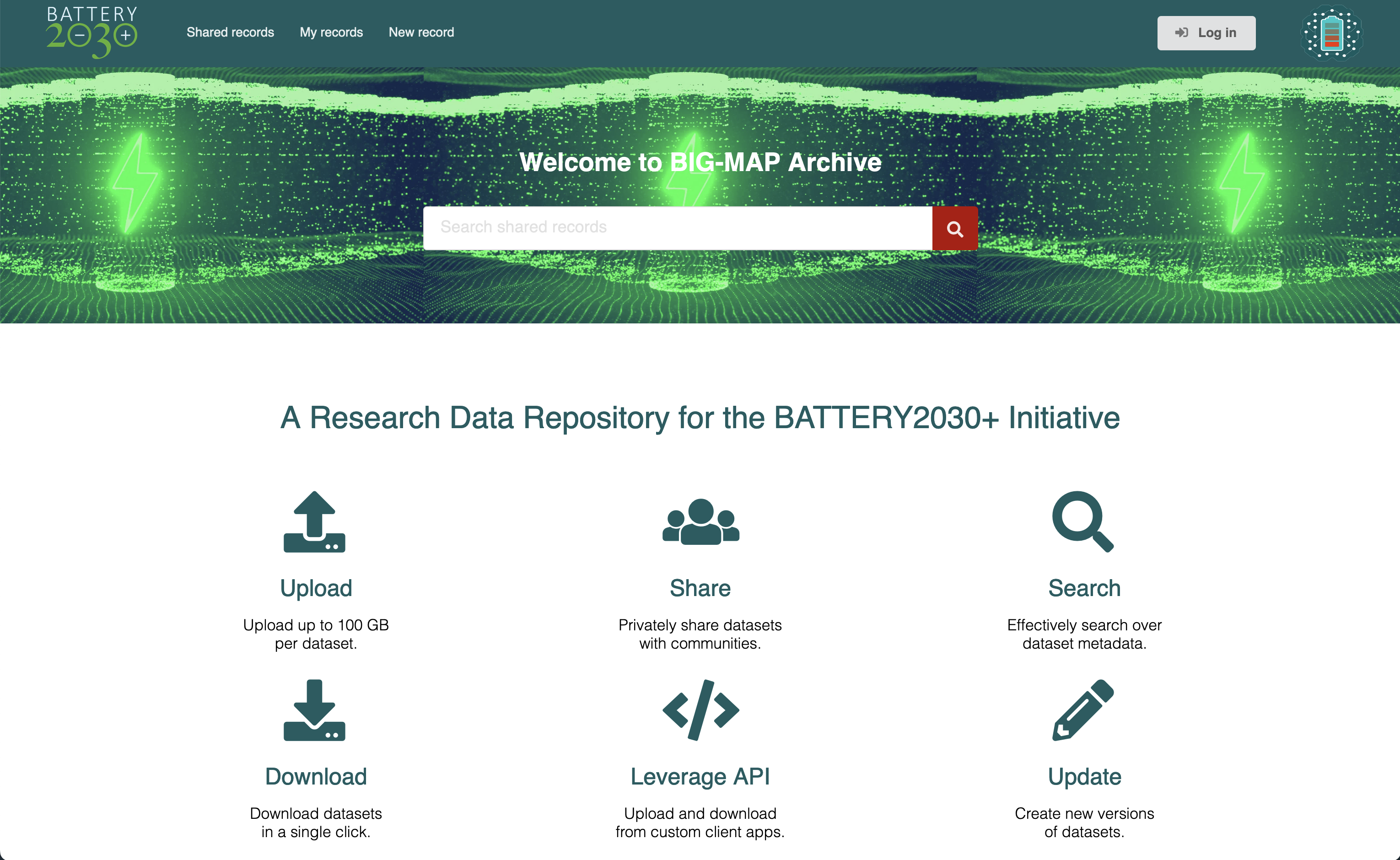}}
\caption{\label{fig:bma}Landing page of the BIG-MAP Archive, a domain-specific and private repository. Access to the repository is restricted to members of the BATTERY 2030+ initiative.}
\end{figure}

The BIG-MAP Archive (Fig.~\ref{fig:bma}) is a web-based, disciplinary and private data repository that enables storage and sharing of research data generated by the BATTERY 2030+ consortium \cite{battery2030}. It has been developed with the concepts described above in mind, promoting collaboration by sharing data and information within the consortium while preserving confidentiality.
Initially built exclusively for members of the BIG-MAP project, it has been opened to all projects of the consortium. It is specifically designed to meet the needs of the consortium, offering a solution for managing sensitive data while fostering collaboration among the members of the project covering a range of different modalities\cite{Vegge2021}. Metadata is structured and formatted to enhance findability and reusability in accordance with the FAIR principles \cite{fair}, making the records ready to be published in public repositories.
The repository hosts today more than 260 records, and has more than 450 registered users, for a total of about 1.5 TB of data uploaded. The repository is built upon InvenioRDM \cite{inveniordm}, a research data management (RDM) repository platform based on the Invenio Framework and developed at CERN. It leverages some of the features of InvenioRDM, such as the communities functionality and the tokenized link sharing, to meet the unique needs of a large consortium seeking a FAIR, secure, and flexible way for data sharing.  

\begin{figure}[tb]
\centering
\includegraphics[width=\linewidth]{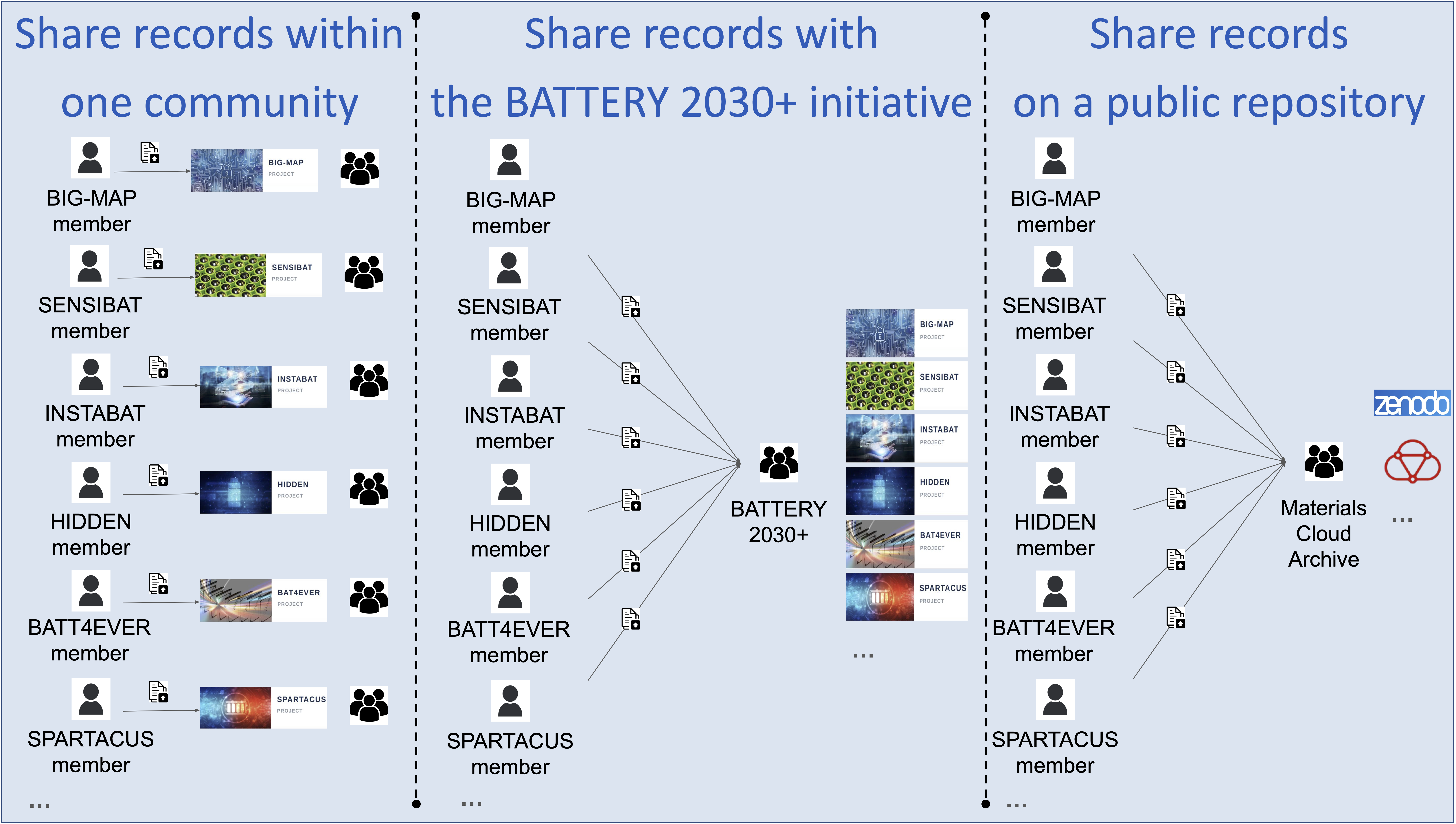}
\caption{\label{fig:sharing_levels} Users can choose to share their records either with members of their own project (called a ``community'' in the language of InvenioRDM) or with all members of the BATTERY 2030+ initiative. 
Sharing records in the BIG-MAP Archive requires authors to curate their data and prepare metadata. This makes it easier for them to publish their results in public repositories such as Zenodo\cite{zenodo} and Materials Cloud Archive\cite{MC} thereafter.}
\end{figure}

The BIG-MAP Archive provides users with a tailored data sharing solution, offering three levels of openness: sharing with members of a specific project, sharing with all members of the consortium, and possibly making data publicly available worldwide (Fig.~\ref{fig:sharing_levels}) by easily and automatically transferring data and metadata to open repositories.
Moreover, thanks to tokenized links, the user keeps full control over what data is shared and with whom, thus providing a granular level of access control and ensuring that sensitive information is only accessed by authorized individuals.
As the application is written using open source software, with the code freely available on GitHub \cite{bma_github}, members of the consortium can inspect and verify the repository software functionalities, ensuring transparency in the way security is implemented, and data is stored and handled. 
Overall, the BIG-MAP Archive gives the consortium full control over how data is shared, ensuring that the owner of the data supervises its access and thus promoting trust within the collaboration.

\section{Technologies and main features of the repository}
The BIG-MAP Archive code is open-source, enabling other consortia to freely access, modify, and redistribute the code \cite{bma_github}. It can thus be considered as a template for other consortia looking to enhance their own data management and sharing capabilities. Institutions can build upon the existing work rather than starting from scratch, and its functionalities and flexibility for customization can be tailored to the unique needs of different research initiatives.

The BIG-MAP Archive is built using InvenioRDM \cite{inveniordm}, a research data management repository platform based on the Invenio Framework. InvenioRDM has a modern web architecture, it relies on open source software and systems, such as OpenSearch, PostgreSQL, Python3/Flask, and has a frontend written in React.

The repository was originally hosted on virtual machines at the Swiss National Supercomputing Centre (CSCS) \cite{cscs}, but has now been moved to the Technical University of Denmark (DTU) \cite{dtu} and the long-term and large-scale Pioneer Center for Accelerating P2X Materials Discovery (CAPeX)\cite{capex} to ensure continuity and long-term sustainability. At DTU, the application is served directly from DTU-managed servers, and all associated research data are stored within DTU's storage infrastructure with daily automated backups to ensure reliability and long-term preservation.

InvenioRDM's robust infrastructure and architecture \cite{invenio_architecture} are designed to scale seamlessly, enabling the platform to handle an increase in traffic or data without compromising performance. 
This is achieved thanks to techniques such as load balancing, caching, and distributed computing which, if needed, enable the application to scale horizontally (over several servers) or vertically (by increasing the power of each server) to meet growing demand.

InvenioRDM uses an hybrid approach for performing read/write operations. 
It stores records in a relational database (PostgreSQL \cite{postgres}) that is used as the primary data source, and keeps a secondary copy indexed in a NoSQL system (OpenSearch \cite{opensearch}) for faster reads. Search queries are sent to the OpenSearch search engine that is JSON-based, highly scalable and has powerful search and aggregation capabilities. 

Metadata is stored according to JSON schemas that provide a standardized way to define its structure and format. This facilitates data integration with other systems and allows for fast and accurate data retrieval when searching for information. Additionally, data retrieval is enhanced through the use of keywords, badges, and filters. These features help users to quickly and efficiently locate the information they need by narrowing search results, visually highlighting relevant information, and further refining the search by applying specific criteria.

In addition to the search functionalities, the repository's intuitive and user-friendly interface offers a range of features to help users to efficiently manage their data. These features include a personal workspace for organized access to the user's records, and an upload form for editing metadata and uploading data.

The upload form has a clean layout, designed to be intuitive and easy to navigate and to provide a step-by-step process guiding users through the upload process. It includes fields to add metadata, such as title, description, and keywords to organize and make the data more easily searchable. 
Moreover, to ensure that authors are properly cited and credited, the upload form uses standardized identifiers. Users can link the authors of a record to their ORCID (Open Researcher and Contributor ID) and affiliations to ROR (Research Organization Registry) IDs, providing a way to accurately identify and disambiguate authors and research institutions.

The upload form has a wide range of licenses to assist users in making informed decisions about how their work is shared and reused. The available licenses include popular and standardized licenses such as Creative Commons (CC), GNU General Public or MIT, as well as the repository's BATTERY 2030+ custom license\cite{battery2030license}. 

Files can be easily uploaded by either dragging and dropping them into the designated box of the upload form or by selecting them individually. A convenient option for uploading multiple files simultaneously is also available. Moreover, a progress bar keeps the user informed of the status of the upload process, while also displaying the total size of the uploaded files. The storage limit per record is 100GB, but it can be increased if necessary.

After finalizing a record and making it available to the other users of the Archive, users can update its metadata, but not its files. To add, remove, or replace files of a shared record, it is instead possible to create new versions, with each version linked and accessible from any other version of the same record. 
Record metadata can be exported in a variety of formats, including JSON, JSON-LD, DataCite, and Dublin Core. This allows to easily integrate records in a range of other systems and tools, such as other data repositories, bibliographic management software, and research information systems.

The first release of the BIG-MAP Archive was developed using InvenioRDM version 9. This release was intended exclusively for BIG-MAP members, and the concept of communities had not yet been included. With the subsequent update to InvenioRDM version 12, the Archive leveraged and adapted the community feature provided by this new version, enabling access for other projects. The Archive is now accessed by members of nine projects within the BATTERY 2030+ consortium.

\subsection{Enabling selective data sharing: community sharing and tokenized links}
One of the key features of the BIG-MAP Archive is its ability to manage access to records through a permission system with granular control over who can access data, thereby ensuring that sensitive information is available only to those for whom it is intended.

The repository is restricted to registered users only, meaning that only people with an account and a confirmed email address can log in and access the information stored in it. Accounts are created for individuals who are directly involved in one or more of the BATTERY 2030+ projects. The list of the members of each project is managed by its project board of directors or the persons appointed by them for user management, who are responsible of providing access only to authorized individuals, and revoking access to people leaving the project.

In the repository, several communities are created: one for each BATTERY 2030+ project, and one additional BATTERY 2030+ community encompassing all projects thst are members of the initiative. 
When a user account is created, it is attributed to all project communities of which the user is a member, as well as the BATTERY 2030+ community.
Since, as discussed below, any record is shared within a community, this makes it possible for users to engage with both the project(s) they are involved in (encouraging project-specific discussions, collaboration, and knowledge sharing) and the broader BATTERY 2030+ community (facilitating cross-project exchanges).

\begin{figure}[tb]
\centering
\includegraphics[width=\linewidth]{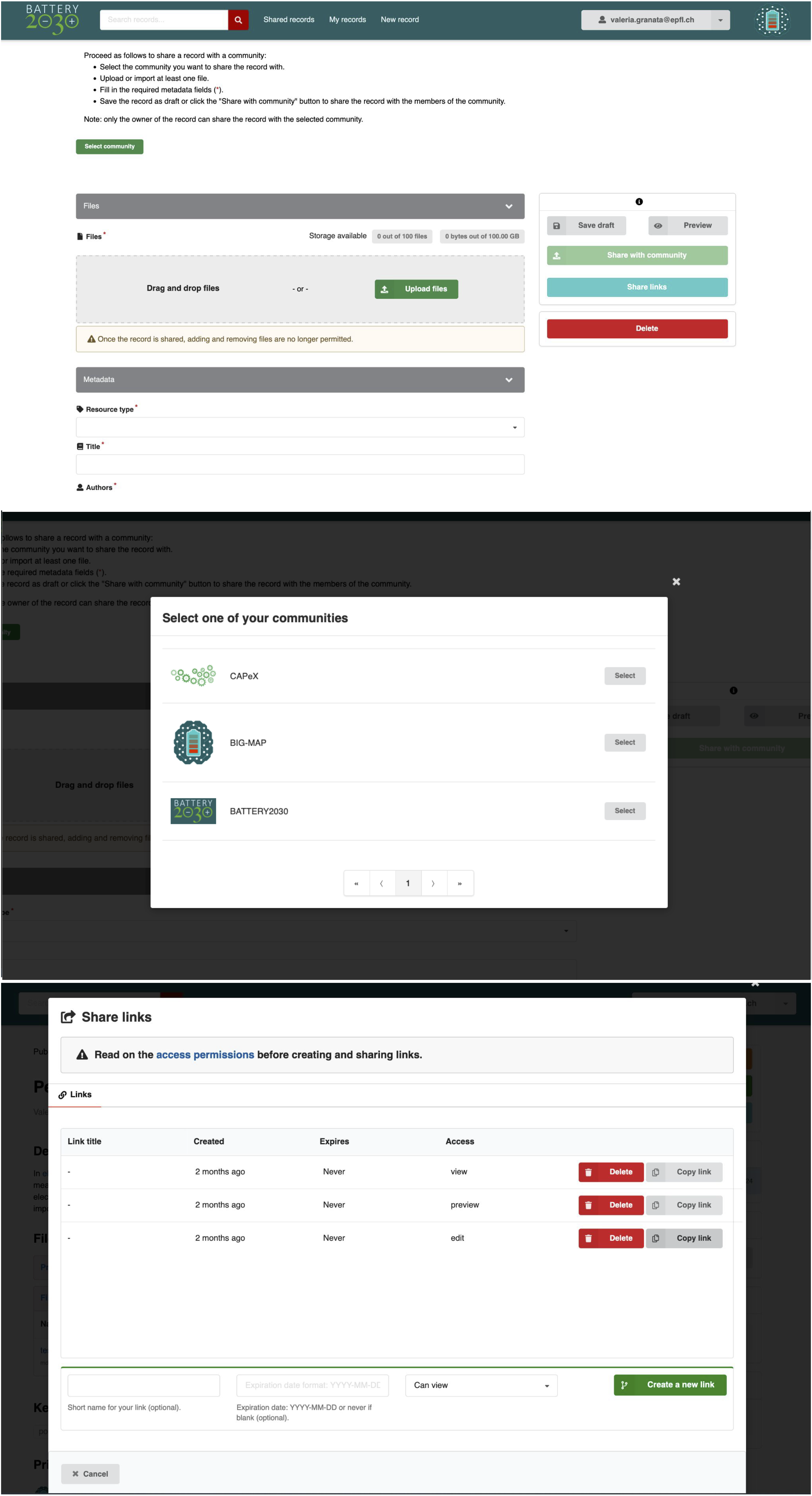}
\caption{\label{fig:sharing} In addition to sharing with communities, users can grant more restricted access permissions through share links to provide specific access (view-only, edit, ...) to selected users with whom the link is shared.}
\end{figure}
Multiple levels of access and granularity can be assigned to a record permissions. This granularity provides fine-tuned control over what actions users can perform on records, such as read-only, edit, delete, or create new versions. 
The most restrictive access level is when the record is saved as draft, at which point the record is kept private to the owner (i.e., the creator of the record), meaning that it can only be accessed and modified by the owner. The next level is sharing the record with the user project's community. From the upload form, the user can select the community with which to share the record. 
Once shared, the record becomes accessible to any member of the community, who may read the metadata and download its files, but not to other communities. Edit permissions remain restricted instead to the record owner.
Finally, by sharing the record with the BATTERY 2030+ community, members of any community in the consortium can access (i.e., read) the record.
The record owner retains control over the permissions granted to the record even after it has been shared with a community. This ensures that the original creators have ultimate authority over their own content and can make any necessary changes without interfering with others.

In some cases, users might need to grant access to records for review, feedback, or approval from colleagues before sharing them with a community. Moreover, the owner may want to grant access to a record already shared with a community to selected colleagues from another community.
To address these situations and allow users to manage access to records in a more granular way beyond just sharing with communities, users can generate ``share links'' (Fig.~\ref{fig:sharing}).
Share links can be generated by the record owner with a specific permission level (view-only or edit) and can then be easily distributed to other users, similarly to the share feature on platforms like Google Docs, Dropbox and the like. 
In particular, share links to view records can be distributed to members of any community and even to external individuals who are otherwise not authorized to access the repository. E.g., when submitting a scientific paper describing a dataset in the Archive, this feature is very valuable to give read-only access to the data (potentially still private during the review phase) to the anonymous referees. Conversely, share links to edit records are limited to members of the same community as the record owner, thereby safeguarding against unauthorized tampering of collaborative work.

\subsection{Secure and anonymous usage statistics}
Usage statistics give valuable insights into how users interact with the repository. The repository offers metrics such as the number of views and file downloads, revealing how frequently records are accessed and which files are most downloaded. This information can be useful for users to understand the reusability of their work.
Moreover, usage statistics are a valuable instrument for project managers to showcase the success and impact of their projects and to make informed decisions about future collaborations and resource allocation. Indeed, they provide concrete evidence of the effectiveness of the project, demonstrate collaboration and overall usability of their research findings, and enhance the credibility of grant applications by providing proof of the impact of previous project results. 
Furthermore, these statistics are also useful for repository managers to increase discoverability and enhance user experience by identifying areas for improvement in metadata management.

\begin{figure}[tb]
\centering
\fbox{\includegraphics[width=\linewidth]{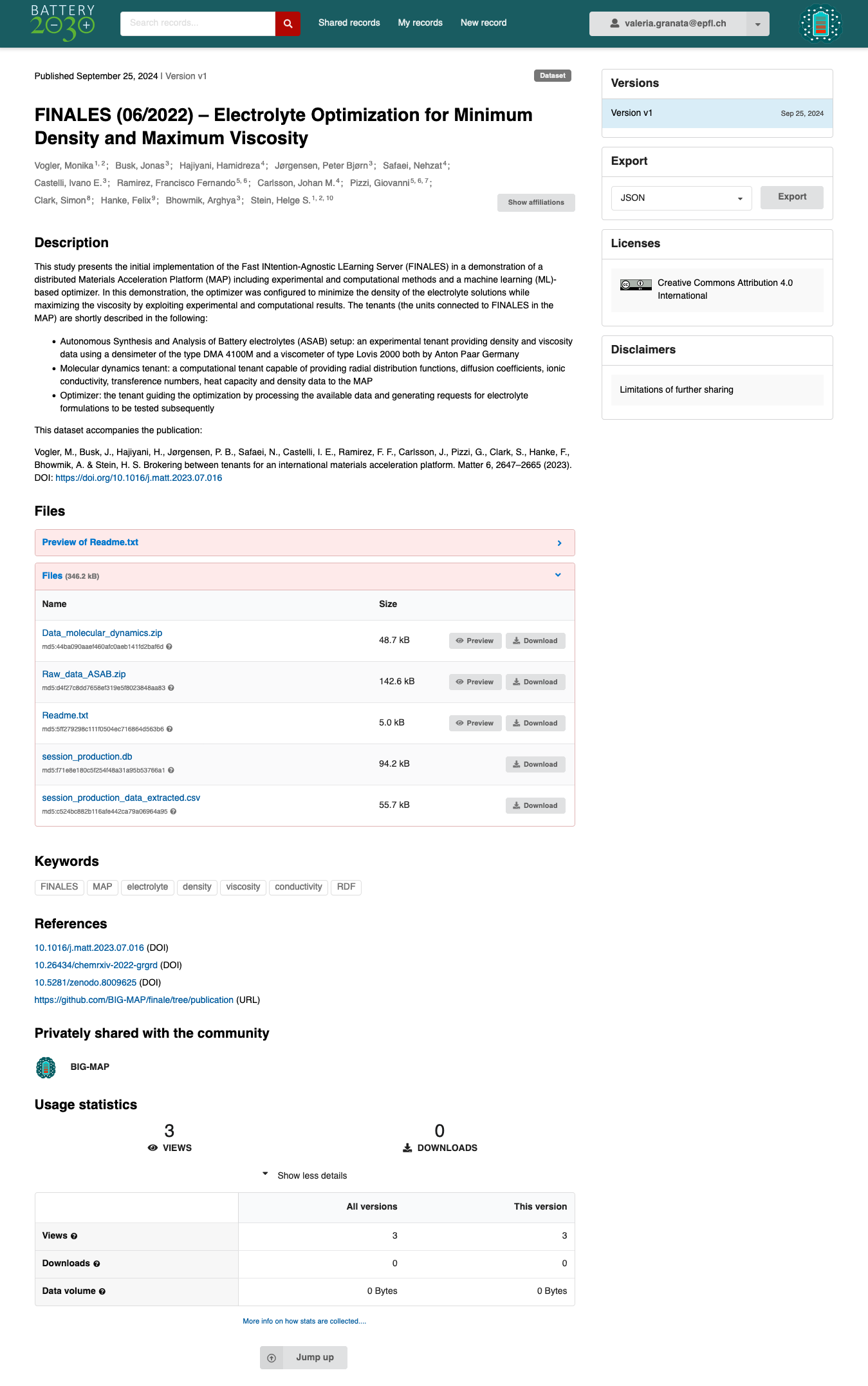}}
\caption{\label{fig:record}An example of a record page. The BIG-MAP Archive tracks usage statistics for each record and its versions, and the individual and cumulative statistics are displayed on each record (see bottom part of the figure).}
\end{figure}

In the BIG-MAP Archive, we use InvenioRDM's usage tracking functionality for collecting statistics. The usage tracking is fully anonymized and is done on the server side. Usage is collected for each record and for each version of the record (Fig.~\ref{fig:record}). The tracked information for view and download events are an anonymized visitor ID, the country of origin of the request (based on the user IP address) and the referrer domain.
The anonymized visitor ID is used to count only unique views and downloads. It is generated by combining the user personal identifier (e.g. the user ID in the repository, the user session ID, or the user IP address) with a salt (a random text value) and applying a one-way cryptographic hash function to scramble the data. The salt is regenerated every 24 hours.

Usage tracking complies with the COUNTER Code of Practice \cite{usage_statistics_counter1} and the Code of Practice for Research Data Usage Metrics \cite{usage_statistics_counter2}, making the repository COUNTER-compliant. 

\begin{comment}
    - More information can be extracted if needed on the time of access and location
\end{comment}

\subsection{REST APIs for data exchange and integration with external applications}
RESTful APIs offer a versatile and scalable approach to facilitate communication between software applications, enabling the exchange of data in a structured and standardized manner.

InvenioRDM's REST APIs are a reliable and efficient way to exchange data with external applications, enabling seamless interactions with the repository. To perform actions, such as creating, updating, sharing records with a community, or creating new versions of a record, users can bypass the graphical interface and interact directly with the repository's backend via scripts. API scripts can be run manually or scheduled to run at regular intervals using tools like cron jobs or task schedulers. This is particularly useful for systematic data retrieval and update, allowing for automated workflows and data synchronization across systems. 

While REST APIs offer a powerful tool for exchanging data, users may need at least a basic knowledge of scripting languages to effectively use them, e.g. Python, JavaScript, or Bash, among others. This is because APIs require making HTTP requests and manipulate data in a structured format.
To make REST APIs accessible to any user and eliminate the need for programming expertise, a command-line API Client has been developed. The BIG-MAP Archive API Client \cite{bma_api} provides simple commands that allows users to easily interact with the repository and execute API calls without having to write code. The Client is distributed as a Python package that can be downloaded and installed from PyPI \cite{bma_api_pyPI}. To authenticate to the repository, users simply need to provide their authentication Bearer token, a short string that can be generated once by the user on the web interface (from the account settings page).

As an example application, the BIG-MAP Archive API Client was successfully used for data exchange with the FINALES brokering software system. FINALES (Fast INtention-Agnostic LEarning Server)\cite{finales,finales2} is a software for research orchestration of multiple tenants, i.e., optimizers, experiments, simulations, and databases. One of the API Client commands allows to back up the FINALES server's SQLite database into the BIG-MAP Archive, and the backed-up data is instantly shared on the repository. During an autonomous research campaign, when new measurements or simulations are performed, they are integrated in the FINALES platform. A specific tenant can be scheduled to periodically trigger the API Client and automatically upload the most recent version of the database to the repository, so that all data becomes immediately available to all other project users. This was demonstrated in the battery optimization campaign of Ref.~\onlinecite{finales2}. Complying with FAIR principles, the data generated during two autonomous FINALES research campaigns\cite{finales,finales2} were made publicly available at the time of publication on the Materials Cloud Archive\cite{MC} at Refs.~\onlinecite{mca_finales1, mca_finales2}.

\section{Future objectives and outlook}
As part of the continued growth of the Archive, three main updates are foreseen: setting up an authentication service using Keycloak, automating the third and final level of openness by streamlining publishing records on a public repository, and integrating new data standards via the BattINFO Ontology. To ensure resources for these developments and longer-term maintainability, the Archive is now managed and hosted by DTU and the Pioneer Center CAPeX, as discussed below.

\subsection{Authentication and users management with Keycloak}
The creation and management of user accounts of the BIG-MAP Archive is currently a task of the repository administrators, who follow the guidelines of the project managers for updating the list of project members.
To facilitate and automate the management of user accounts in the repository and allow for separation of concerns (functionalities), a centralized service based on Keycloak \cite{keycloak} will be implemented. 
Keycloak is an open-source identity and access management (IAM) solution, with a robust and scalable authentication and authorization system. It is based on standard protocols and provides support for OpenID Connect, OAuth 2.0, and SAML.
Keycloak's authorization framework is built upon the concepts of roles, permissions, scopes, and policies, which collectively enable a granular control over user access. InvenioRDM supports integration with external authentication methods, including OAuth, and offers out-of-the-box integration with Keycloak. 

Once the Keycloak server is setup and the expected roles and permissions are defined, user credentials and other personal data will be securely stored in the Keycloak database. 
Keycloak uses end-to-end encryption and secure protocols to protect user data and ensure the integrity of authentication and authorization transactions.
Additionally, Keycloak has Single Sign-On (SSO) functionality.
If other applications in addition to the BIG-MAP Archive are configured to access the same Keycloak provider, the SSO feature allows users to access all of them with a single set of login credentials, without the need to log in to each application individually.

This approach lets the BIG-MAP Archive application focus on its primary business functionalities, while delegating all user authentication and authorization processes to Keycloak. This also allows for the separation of roles and tasks in the project between repository administrators, focusing on the management of the BIG-MAP Archive, and user managers, who can keep the user list updated directly only Keycloak.

\subsection{Sharing data on a public repository}
\begin{figure}[tb]
\centering
\fbox{\includegraphics[width=0.95\linewidth]{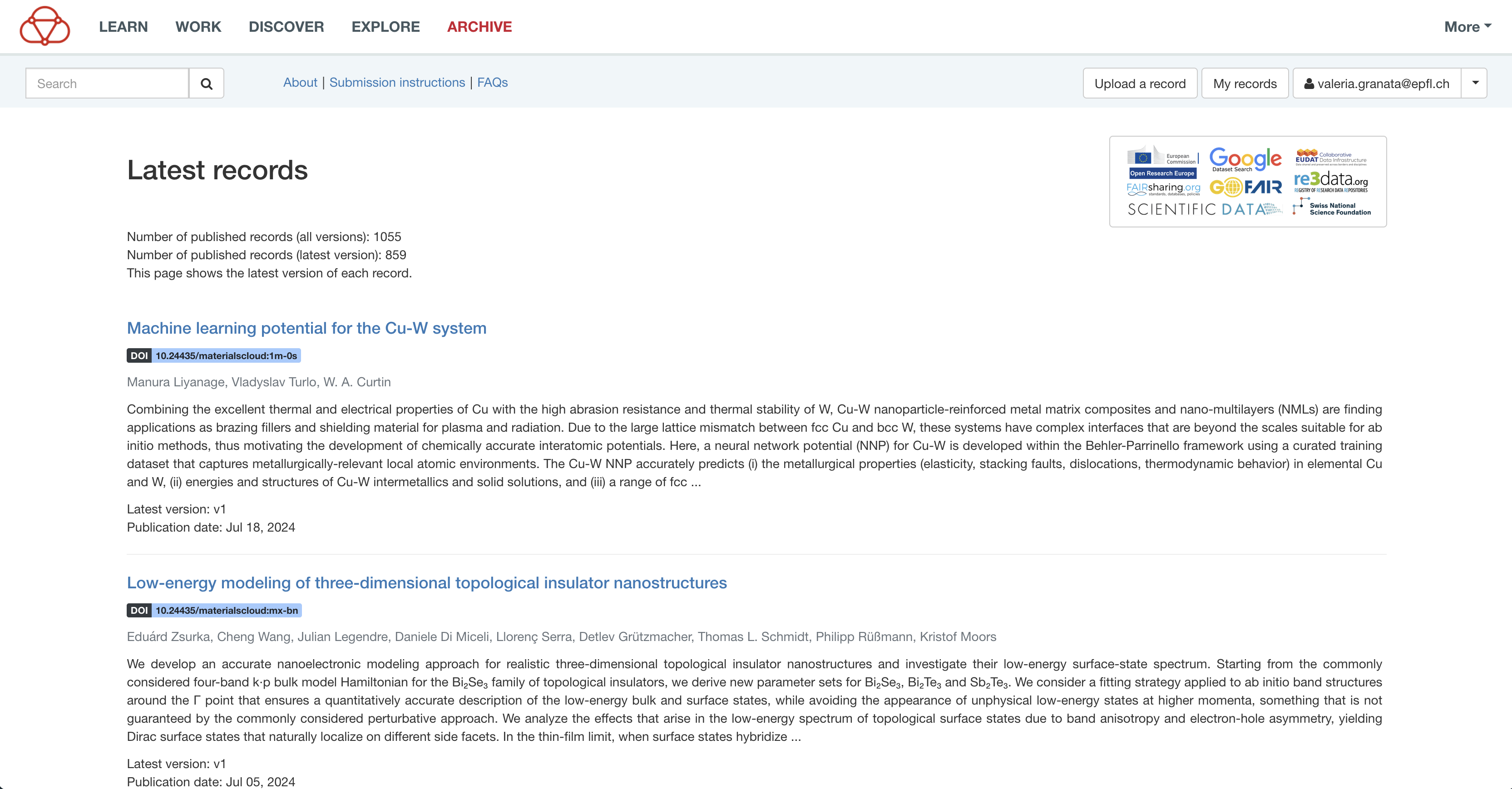}}
\caption{\label{fig:mca} The Materials Cloud Archive is a moderated and public repository for computational materials science. With a single click, users will be able to publish records shared on the BIG-MAP Archive directly to the Materials Cloud Archive.}
\end{figure}

The BIG-MAP Archive currently offers users the two options of sharing records with their project members or with the entire BATTERY 2030+ initiative. A third option will be developed for publishing records to a public data repository, the Materials Cloud Archive\cite{MC,mca} (Fig.~\ref{fig:mca}), a moderated, open repository for research data relevant to computational materials science. The repository provides Digital Object Identifiers (DOIs) to each published entry, ensuring that they can be easily found and referenced.

In the planned upgrade, users will be able to automatically transfer records created and shared within the BIG-MAP Archive to the Materials Cloud Archive, where they will then be reviewed by moderators and published, once approved.
To automate the process of uploading records from the BIG-MAP Archive to the Materials Cloud Archive, a specific CLI command will initially be implemented in the BIG-MAP Archive API Client. This command will copy the metadata and files from a shared record on the BIG-MAP Archive to a draft record on the Materials Cloud Archive. Once created, the user can modify the draft as needed before submitting it for review.
To ensure a secure authentication, users must also have an account on the Materials Cloud Archive and provide their credentials or token when running the CLI command. However, we note that this extra step will no longer be necessary if both Archives will implement a Keycloak SSO functionality and use the same authentication server. In this case, the same authentication can be reused, eliminating the need to log in to each Archive individually.

In a second phase, the data transfer between the two Archives might be made available directly via the BIG-MAP Archive user interface, where users will initiate the transfer with just a click of a button, further streamlining the process and reducing the need for manual commands.

\subsection{Enabling FAIR-by-Design data for autonomous platforms}
One of the primary objectives of the BIG-MAP Archive is to prepare data for reuse and reprocessing by other systems. For effective integration into other systems, data needs to be structured and labeled using a common standardized vocabulary for concepts. A mapping should be provided to format the data into a semantic data model that matches data fields with their corresponding meaning, ensuring data is machine-readable and interoperable.
An example of such mapping is the BattINFO ontology \cite{battinfo}, designed to support the development of computational tools and the deployment of interoperable data in the BIG-MAP project and beyond.
The BattINFO ontology is specific to the battery domain and describes thousands of concepts related to electrochemistry and batteries, as well as the relations between these concepts. By giving meaning to data, it creates links to other pieces of data and allows for machine reasoning according to that meaning.

An example of a system generating BattINFO-compliant metadata is the FINALES autonomous framework \cite{finales,finales2} described earlier. 
All communication with FINALES must happen via standardized data schemas, and these can also be enriched with their ontological annotations.
When data and metadata are stored in the BIG-MAP Archive as described earlier, also semantic annotations
can be uploaded to the Archive in standard formats, such as JSON-LD.
The Archive can then be adapted to highlight the availability of these semantic annotations in an entry and visually display them.
Notably, this makes it easier to both understand the data and search for specific information, making entries easily findable via semantic searches that could be provided by custom ontology-aware search platforms.

By uploading data along with their semantic annotations to the Archive, the FAIR principles will be taken into account at every stage of a project, ensuring that data is created, shared and maintained from the very beginning in a manner that enhances its usability and value. As a result, this approach embraces a FAIR-by-Design methodology by integrating the FAIR principles of data management into the design and development of projects such as FINALES. 

\subsection{Future directions and scale-up}
As mentioned earlier, the BIG-MAP Archive has been moved to ensure long-term sustainability and continuity of the service, and is now maintained by DTU and the CAPeX center. The original URL of the Archive (\url{https://archive.big-map.eu/}) now automatically redirects to the new hosting location at \url{https://archive-capex.energy.dtu.dk}.
Being based on the robust InvenioRDM technology, the Archive is fully scalable to include other projects within and beyond BATTERY 2030+, such as Salamander, UltraBat, and BatCAT, and can be further extended to areas beyond batteries, covering the upcoming Materials Commons initiative to be launched by the European Commission in 2026 (see e.g., \url{https://materialscommons4.eu}). Moreover, the Archive can be made interoperable via its API with other data storage facilities, such as the Horizon Europe project DECODE\cite{decode} on developing green hydrogen technologies.

The archive is further intended to play a central role in the new pan-European RAISE program\cite{raise} that was launched by the European Commission in Copenhagen on November 4th, 2025, as part of the AI for Science Summit\cite{ais25}. Here, a specific call to action was made from the AI for Materials Science community for a RAISE-AM: ``A Pan-European RAISE laboratory for Advanced Materials'' that was pitched as the outcome of the AI in Materials Science workshop. 

Initiatives like Materials Commons and RAISE-AM would be strongly dependent on repositories like the BIG-MAP Archive to support access to large-scale multi-sourced and interoperable datasets, e.g., for training of European-based foundation models, such as the MACE-MP-0 foundation models for materials chemistry\cite{Batatia2025}. For this purpose, a central Mat4EU hub is envisaged to link federated archives and data spaces for advanced materials and related initiatives.   

\section{Discussion and conclusions}

The BIG-MAP Archive addresses the unique challenges of data sharing within large research consortia, particularly those involving sensitive information. By providing a secure and user-friendly platform tailored to the needs of the BATTERY 2030+ initiative, the repository facilitates collaboration while ensuring confidentiality through controlled access. The implementation of advanced features, such as tokenized sharing links and fine-grained permission settings, empowers researchers to manage their data effectively while promoting collaboration among members of diverse projects.

The technologies used to develop the repository, built on the InvenioRDM platform, enhance its scalability and adaptability, ensuring that it can accommodate the evolving demands of its users. The commitment to the FAIR principles further ensures that the data is not only preserved but also remains discoverable and reusable, allowing for integration with other applications and tools.

Looking ahead, the planned enhancements, including the integration of Keycloak for improved users management, the ability to publish records in public repositories such as the Materials Cloud Archive, and the integration of the BattINFO ontology, will significantly improve the repository's capabilities. These enhancements aim to create an even more robust system for scientific collaboration, facilitating seamless data exchange.

The BIG-MAP Archive sets a valuable model example for other consortia and research initiatives. It has proven to be an important tool for the BATTERY 2030+ initiative, fulfilling essential roles in data management and sharing\cite{Castelli2021}. Additionally, its open-source code serves as a template for other consortia seeking to develop their own data repositories.

\section{Data availability statement}
No new data were generated or analyzed as part of this paper. However, we stress that the BIG-MAP Archive itself is a repository to host data. Access to individual records is granted to members of the communities with which the records have been shared.
The source code for the BIG-MAP Archive is openly available on GitHub at \url{https://github.com/team-capex/big-map-archive} and archived on Zenodo at DOI: \url{10.5281/zenodo.17939528}.

\section{Acknowledgements}
This project received funding from the European Union's Horizon 2020 research and innovation program under grant agreement no. 957189 (BIG-MAP). The authors acknowledge BATTERY2030PLUS, UltraBat, and BatCAT, funded by the European Union's Horizon 2020 research and innovation program under grant agreement no. 957213, 101103873, and 101137725, and the Pioneer Center for Accelerating P2X Materials Discovery (CAPeX), DNRF grant number P3.
VG, XW, NM and GP acknowledge financial support by the NCCR MARVEL, a National Centre of Competence in Research, funded by the Swiss
 National Science Foundation (grant number 205602). 
We acknowledge support by the Open Research Data Program of the ETH Board (project ``PREMISE'': Open and Reproducible Materials Science Research, and project ``API-03 IntER'': Interoperability between the ETH Domain
Repositories).

%\printbibliography
%

\end{document}